\documentclass[useAMS,usenatbib]{mn2e}
\usepackage{graphicx} 
\usepackage{pslatex}
\usepackage{bm}

\title[Strong lenses and time delays in nonuniform universe]
{The influence of the matter 
along the line of sight and in the lens environment on the strong 
gravitational lensing}
\author[M. Jaroszynski and Z. Kostrzewa-Rutkowska]
{M. Jaroszynski$^{1}$\thanks{E-mail:
mj@astrouw.edu.pl (MJ); zkostrze@astrouw.edu.pl (ZKR)} 
and Z. Kostrzewa-Rutkowska$^{1}$\\
$^{1}$University of Warsaw Observatory, Al. Ujazdowskie 4, 00-478 Warsaw,
Poland}

\begin{document}

\date{Accepted; Received ;}

\pagerange{\pageref{1}--\pageref{10}} \pubyear{2014}

\maketitle

\label{firstpage}

\begin{abstract}
We investigate the influence of the matter along the line of sight and in the 
lens environment on the image configurations, relative time delays, and the 
resulting models of strong gravitational lensing. The distribution 
of matter in space and properties of gravitationally bound haloes
are based on the Millennium Simulation. In our numerical experiments we
consider isolated lens in a uniform universe model and the same lens 
surrounded by close neighbours and/or objects close to the line of sight
which gives four different descriptions of the light propagation. We compare 
the results of the lens modeling which neglects effects of the environment 
and line of sight, when applied to image configurations resulting 
from approaches  partially or fully taking into account these effects.
We show that for a source at the redshift $z\approx 2$ the effects are 
indeed important and may prevent successful 
fitting  of lens models in a substantial part of simulated image 
configurations, especially when the relative time delays are taken into 
account. To have good constraints on the models we limit ourselves to 
configurations of four images. We consider eighty lenses and large number 
of source positions in each case. The influence of the lens neighbourhood 
and the line of sight introduces the spread into the fitted values of 
the deflection angles which translates into the spread in the lens 
velocity dispersion $\sim$ 4 per cent. 
Similarly for the lens axis ratio we get the spread of $\sim$ 10 per cent
and for the Hubble's constant of $\sim$ 6 per cent.
When averaged over all lenses and all image configurations considered, 
the median fitted values of the parameters (including the Hubble's constant) 
do not differ more than 1 per cent from their values used in simulations.

\end{abstract}


\section{Introduction}

The strong gravitational lensing has many astrophysical and cosmological 
applications (see e.g. \citealp{K06} and references therein for a review). 
The qualitative understanding of the majority of multi-image configurations
attributed to strong lensing can usually be based on a model of single 
galaxy-lens in a uniform universe. 
The quantitative description requires more complicated 
models, starting with so called {\it external shear} \citep{ChR84}, 
invoking a galaxy cluster \citep{Y81}, or taking into account another 
galaxy in the lens vicinity (\citealp{Koo98}, \citealp{McL09}).
The influence of the mass distribution in the lens vicinity and along
the line of sight has been investigated by many authors 
(\citealp{KA88}, \citealp{KKS97}, \citealp{BK96}, \citealp{ChKK03},
\citealp{WBO04}, \citealp{WBO05}, and \citealp{AN11} to cite few).

Photometric survey of several strong lens surroundings \citep{Will06} 
shows that many of them lie in poor groups of galaxies and that other 
groups near the line of sight are not uncommon. Spectroscopic observations 
(\citealp{Mom06}, \citealp{Aug07}) give the distribution of the galaxies 
along the line of sight and allow more detailed study of their grouping 
and influence on strong lensing in several cases. The inclusion of the 
directly observed objects around the lens in modeling greatly improves 
the quality of fits. 

\citet{KZ04} investigate theoretically the problem of the main galaxy 
close neighbours constructing a poor group of galaxies. They check the 
image configurations corresponding to various source positions behind 
the group, different group members playing the role of the main lens, and 
others playing the role of the environment. They thoroughly analyze 
the influence of lens environment on the values of the fitted parameters.
They show that by neglecting the objects around the lens one introduces 
bias to the fitted parameter values, which plays the role of a systematic 
error. 

In this paper we continue our investigation of the environmental and line
of sight effects which influence the action of strong gravitational lenses
using the results of the Millennium Simulation \citep{Spr05} from its
online database \citep{ls06}.
We basically follow the approach of \citet{JK12} (hereafter Paper I) 
including also the time delays in our considerations.
We attempt to quantify 
the influence of matter in the strong lens environment ({\it ENV})
and along the  line of sight ({\it LOS}) on the results of modeling 
multiple image configurations with measured time delays.  
We simulate the propagation of light in four different ways. In the 
most simplified approach we include only the isolated strong lens in 
a uniform universe model. Other approaches include the lens environment,
or the matter along the line of sight, or both. Using each of the 
approaches we simulate many multiple image configurations, and attempt 
to fit them with the same kind of simplified model. The rate of failure 
(i.e. the fraction of unsuccessful fits in each approach) 
measures the influence
of the environment and the line of sight (or each of them separately) 
on the strong lens. The differences between the fitted values of model 
parameters and the parameters used in simulations give 
the estimate of the systematic errors introduced by the environment 
and the line of sight. Our goal is the comparison of various effects 
related to light propagation, not the improvement of strong lens modeling. 
 
In Sec.~2 we describe our approaches to light propagation. 
Sec.~3  presents tools used to compare different models and the results of such
comparison. Discussion and conclusions follow in Sec.~4.

\section[]{Model of the light propagation}

\subsection{Ray deflections and time delays}

The multiplane approach to gravitational lensing (e.g. \citealp{SW88};
\citealp{SS92}) using the results of the Millennium Simulation \citep{Spr05}
and the non-singular isothermal ellipsoids (NSIE) as models for individual 
halos (\citealp{KSB94}; \citealp{K06}) is described in Paper I. 
Here we augment it with the description of relative time delays. 

The evolution of the 
matter distribution is given by the Millennium {\it snapshots} which 
correspond to several discrete epochs with given redshifts $\{z_i\}$.
We assume that for $(z_{i-1}+z_i)/2 \le z \le (z_i+z_{i+1})/2$ 
the Millennium cube of epoch $z_i$ adequately describes matter distribution.
Thus a ray crosses perpendicular layers of matter of defined thickness 
cut from the Millennium cubes of different epochs. 
The cubes are randomly shifted and rotated to avoid 
effects of periodic boundary conditions of the Simulation  \citep*{b10}.   
Since there are several matter layers between the source at 
$z\ge 1$ and the observer, they can be treated as thin, and may be 
represented as surface mass distributions projected into their middle 
planes.

The matter 
content of each cube is described as a {\it background} component representing
matter density averaged on $256^3$ cells plus gravitationally bound haloes
given by \citet{b11} and \citet{b4}. For the background we calculate the 
gravitational force in 3D and then use its component perpendicular to a ray  
to obtain the deflection angle. For ray beams with the small opening 
angles of $\sim 3~\mathrm{arcmin}$ the major influence of each background cube 
is an almost constant deflection angle $\bm{\alpha}_\mathrm{bcg}$ plus its
small variation, which we describe as the action of the background convergence
$\kappa_\mathrm{bcg}$ and shear $\gamma_{1,\mathrm{bcg}}$, 
$\gamma_{2,\mathrm{bcg}}$. (These parameters are defined for each layer 
separately.)

Each projected halo is represented as a difference between two
NSIE distributions with the same characteristic deflection angles $\alpha_0$,
axis ratios $q$, and position angles,
but different values of core radii $r_1 \ll r_2$, which makes its mass finite:
\begin{equation}
\lim_{r\rightarrow\infty}(\bm{\alpha}_1- \bm{\alpha}_2)= 
\alpha_0(r_2-r_1)\frac{\bm{r}}{r^2}
~~\Leftrightarrow ~~ M=\frac{c^2}{4G}\alpha_0(r_2-r_1)
\label{virialmass}
\end{equation}
(compare Paper I). The above formula gives the value of characteristic 
deflection $\alpha_0$ for a halo of given mass and virial radius 
$r_\mathrm{vir}\approx r_2$. (We use $r_1 \ll r_2$, which validates 
the approximation). We consider axis ratios which are distributed within 
$0.5 \le q \le 1$ with maximum probability at $q=0.7$, loosely resembling 
the results of \citet{KY07}. The position angles in the sky are random.
Since the {\it background } contains the whole mass, including mass of haloes,
the latter must be compensated by some negative density distribution. 
We use the constant surface mass circular disk of radius $r_\mathrm{lim}$ 
($r_2 \ll r_\mathrm{lim}$) for this purpose (for details see Paper I).
A compensated halo does not deflect rays outside its $r_\mathrm{lim}$ 
radius, so only a finite number of haloes has to be included in calculations.

A ray coming to the observer from the direction $\bm{\beta}_1$ in the 
synthetic sky, crosses the $N$-th layer ($N \ge 1$) at some position 
$\bm{\beta}_N$ given as \citep{SW88}:
\begin{equation}
\bm{\beta}_N = \bm{\beta}_1 
- \sum_{i=1}^{N-1}~\frac{d_{iN}}{d_{N}}~\bm{\alpha}_i(\bm{\beta}_i)
\end{equation}
where $d_{iN}$ is the angular diameter distance as measured by an observer
at epoch $i$ to the source at epoch $N$, $d_N$ - the angular diameter
distance to the same source measured by a present ($z=0$) observer.
In calculations we use comoving distances, which in a flat cosmological
model simplifies some of the expressions, but we skip these technical 
details here. We also apply more efficient recurrent formula of \citet{SS92}, 
equivalent to the above equation.

Knowing the light path, one can calculate the {\it geometric} part of the 
relative time delay $\Delta t_\mathrm{geom}$ (as compared with the 
propagation time along a null geodesics in a uniform universe model) using 
the formula \citep{SEF92}:
\begin{equation}
c\Delta t_N^\mathrm{geom}(\bm{\beta}_1) = 
\frac{1}{2}~\sum_{i=1}^{N-1}(1+z_i)~\frac{d_{i+1}}{d_id_{i,i+1}}
\left(d_i(\bm{\beta}_{i+1}-\bm{\beta}_i)\right)^2
\end{equation}
where we consider a ray coming to the observer from the direction 
$\bm{\beta}_1$ (which defines its earlier path, so all $\bm{\beta}_i$ are 
known) and the factors $1+z_i$ represent time dilatation.

The deflection in each layer can be calculated as a gradient of the deflection 
potential, which is also a measure of {\it gravitational} time delay 
$\Delta t_\mathrm{grav}$ \citep{SEF92}:
\begin{equation}
\bm{\alpha}_i = -\frac{1}{d_i}\frac{\partial\Psi_i}{\partial\bm{\beta}_i}
~~~~~~c\Delta t_\mathrm{grav}(\bm{\beta}_i) = \Psi_i(\bm{\beta}_i) + C 
\label{gradient}
\end{equation}
The potential is defined up to a constant $C$. The cumulative 
{\it gravitational} time delay after crossing all the layers is:
\begin{equation}
c\Delta t_N^\mathrm{grav}(\bm{\beta}_1)= 
\sum_{i=1}^{N-1}~(1+z_i)\Psi_i(\bm{\beta}_i)
\end{equation}
where again $\bm{\beta}_1$ defines the path and we account for time 
dilatation.
Finally:
\begin{equation}
\Delta t_N(\bm{\beta}_1) 
= \Delta t_N^\mathrm{geom}(\bm{\beta}_1) 
+\Delta t_N^\mathrm{grav}(\bm{\beta}_1) 
\end{equation}
The above expression contains unknown additive constant. Only the 
difference in calculated time delays between two rays can have a clear 
physical meaning.

\subsection{Light propagation in selected solid angles}

Our backward ray shooting covers separate and randomly chosen {\it maps} 
in the synthetic sky $\approx 3\mathrm{arcmin}\times 3\mathrm{arcmin}$ 
in size. To propagate a wide beam of this size we need deflection angles 
in overlapping regions in all layers. 
For this purpose we calculate and store the deflection angles
for the $2048 \times 2048$ evenly spaced positions of interest in each layer.
Similarly we store the {\it gravitational} time delays caused by each layer.
With the help of interpolations one can obtain deflections and time delays 
for any position in any layer.

The deflection angle and the deflection potential are given by 
analytical formulae and for each grid point Eq.~\ref{gradient} holds 
exactly. Interpolations and finite differencing used in the numerical 
approach make this relation approximate. To check the selfconsistency 
of our methods we have numerically calculated 
the rotation of the deflection angle on the grid, for different positions, 
maps, and differencing spacings. The rotation of the deflection field
is present, but its level is fairly low: 
$\left<|\alpha_{x,y}-\alpha_{y,x}|\right>
< 0.01\left<|\alpha_{x,y}+\alpha_{y,x}|\right>$, where the averaging is 
over many positions within a single map. The result holds for maps 
at different redshifts for spacings of $\sim 1~\mathrm{arcsec}$. For
spacings of $\sim 0.1~\mathrm{arcsec}$ the rotation is two times more 
important. 

We choose to calculate the strong lens effects in the nonuniform universe 
model for sources at the redshift $z\approx 2$. We examine all haloes within 
the beam up to this redshift (typically few thousands of them) as lens 
candidates estimating their Einstein ring radii $r_\mathrm{E}$. 
We choose ten lenses with 
the largest Einstein rings as most likely to produce multiple images of 
a randomly positioned source and investigate them in detail.

When investigating multiple image properties we need zoomed maps of
smaller parts of the sky. For better resolution we use finer grids with
deflections interpolated from previous calculations with the help of
bi-cubic spline, so the interpolated deflection derivatives are
continuous. We expect the multiple images to lie within few Einstein 
radii from the halo centre. 
We check for the presence of other haloes inside  the circle of the radius 
$3~r_E$ surrounding a dominating lens. If they are present we enlarge the 
region of interest including $3~r_E$ zones around all companions. Finally 
we repeat backward ray shooting inside a square on the sky overlapping the 
region of interest. The fine grids giving the deflection angles in consecutive
layers encompass still larger areas, so one can follow majority of rays 
deflected off the main region. We keep control of the rays and if some 
of them leave the mapped region, we neglect the related lens in further 
considerations. 

We store the result of the ray shooting as a vector and scalar arrays:
\begin{equation}
\bm{\beta}_N^{kl}=\bm{\beta}_N(\bm{\beta}_1^{kl})~~~~~~
\Delta t_N^{kl}= \Delta t_N(\bm{\beta}_1^{kl})
\end{equation}
where $\bm{\beta}_N^{kl}$ gives the positions in the source plane 
of rays apparently coming from the directions $\bm{\beta}_1^{kl}$ 
on the observer's sky. Similarly $\Delta t_N^{kl}$ denotes the total 
(gravitational plus geometric) time delay along the ray coming from 
the same  direction. Superscripts $k$, $l$ enumerate the rays. 
Calculations for each ray are based on Eq.(2) and Eq. (6) respectively.

\subsection{Numerical experiments: different descriptions of light propagation}

To estimate the importance of the influence of the matter along the 
line of sight ({\it LOS}) and/or 
the matter in the strong lens environment ({\it ENV}) 
on the lensing properties, we repeat the simulations four times. The most 
realistic approach takes into account both {\it LOS} and {\it ENV}. When we 
take into account the action of the strong lens and neglect all other haloes 
belonging to the same layer, the influence of {\it ENV} is lost, but 
{\it LOS} may still be included. In this way we check the effect of 
{\it LOS} alone. Similarly, assuming that rays are deflected only in 
the strong lens layer ($\bm{\alpha}_i \equiv 0$ for other layers) the effect 
of {\it ENV} alone is modeled. Finally using the same strong lens but 
in a uniform universe ({\it UNI}) we have the clear case unaffected by 
neither {\it LOS} nor {\it ENV}, which may be used for comparison.

We use the {\it prismatic transformation} \citep{gfs88} in all layers  
making the deflection at the middle points of all maps zero and transforming 
the time delays accordingly. This implies that removing the background and 
all other haloes in the major lens plane changes the backward ray paths 
beyond the plane only in the second order. Thus the rays in the 
{\it LOS+ENV} and {\it LOS} approaches travel through very similar 
surroundings and the comparison of lens modeling in these cases is sensible.  

\subsection{Synthetic multiple image configurations}

The lens equation giving the dependence $\bm{\beta}_N(\bm{\beta}_1)$ is
stored in the array $\bm{\beta}_N(\bm{\beta}_1^{kl})$ for many positions 
in the image plane and is sufficient for quite accurate interpolations.
Using numerical methods we find the distortion matrix and its determinant
on the same grid, and then the critical lines and caustics. 
Finally we find image positions for many source positions using iterative 
methods. (Compare Paper I). We consider only sources placed inside caustics
since we are interested in multiple image configurations. The source 
positions are evenly spaced inside the region of interest. 
Once the images for a given source position are found, we calculate the 
corresponding flux amplifications and time delays,  
and store all such information for further analysis. 

Different descriptions of light propagation produce different critical 
lines and caustic structures. While the positions on the observer's sky 
can be easily compared independently of light propagation model, the same 
can not be said of source positions. Thus the comparison between image 
configurations corresponding to the same source position but obtained 
with different descriptions of light propagation is meaningless. Only the
statistical properties of image configurations corresponding to given 
propagation model can be compared with one another.

\section{Fitting the simulated strong lenses with simplified models}

\subsection{Simplified fits and their rate of success}

\begin{table*}
 \centering
 \begin{minipage}{116mm}

  \caption{Acceptability of fits - dependence on the model}
  \begin{tabular}{@{}ccccccc@{}}
  \hline
  Shear 	&  no & no & no & yes &  yes & yes\\
  Delays        &  no & yes & yes & no & yes & yes\\
  $H_0$         &  no & no & yes & no & no & yes\\
  model \#      & 1 & 2 & 3 & 4 & 5 & 6 \\
 \hline
 {\it LOS+ENV} & 0.44$\pm$ 0.19 & 0.35$\pm$ 0.19 & 0.39$\pm$0.16 & 
                 0.62$\pm$ 0.17 & 0.52$\pm$ 0.16 & 0.56$\pm$ 0.17\\
 {\it LOS}     & 0.64$\pm$ 0.21 & 0.60$\pm$ 0.20 & 0.63$\pm$ 0.19 & 
                 0.71$\pm$ 0.19 & 0.66$\pm$ 0.18 & 0.67$\pm$ 0.19\\
 {\it ENV}     & 0.65$\pm$ 0.18 & 0.58$\pm$ 0.19 & 0.59$\pm$ 0.18 & 
                 0.85$\pm$ 0.08 & 0.76$\pm$ 0.14 & 0.78$\pm$ 0.11\\
 {\it UNI}     & .995$\pm$ .004 & .994$\pm$ .004 & .995$\pm$ .003 & 
                 .998$\pm$ .001 & .996$\pm$ .002 & .998$\pm$ .001\\
 \hline

\noalign{\vskip3pt}
\multicolumn{7}{p{11.6cm}}{Note: The table shows the dependence of 
the rate of acceptability of fits on the method of treating the light
propagation (see the text for details),  
for models neglecting (``no'') or taking into account (``yes'') 
the external shear, neglecting or taking into account time delays,
keeping fixed (``no'') or modeling (``yes'') the value of the 
Hubble's constant.
The required accuracy is the same for all models
($\sigma_\mathrm{L}=0.1$,  $\sigma_\mathrm{I}=0.003$, $\sigma_{~\Delta m}=0.1$,
$\sigma_{~\Delta t}=0.001~\mathrm{y}$). Models are numbered for further 
reference.}
\end{tabular}
\end{minipage}
\label{success}
\end{table*}

In the case of pure elliptical, 
non-singular lens one would expect (in general) 1, 3, or 5 images.
Our numerical method of finding the images starts from ray-shooting 
on a finite resolution grid with a  small extended source as a target, 
so some images may be missed or unresolved.  
Also the external influence may change the number of images. 
Since configurations with more images probably better constrain the lens, 
we concentrate on cases with five or four images.

To estimate the influence of the galaxies in the lens vicinity and the 
matter along the line of sight on the properties of the strong lensing, 
we attempt to model all our simulated cases of multiple imaging using a 
single lens model in a uniform Universe. The single lens we are using 
in modeling is a non-singular finite isothermal ellipsoid used also in the 
simulations for each of the haloes. When tracing the rays in simulations we 
interpolate all deflection angles from earlier computed arrays. Single 
lens modeling uses analytically calculated deflection angles and their 
derivatives, so it serves also as a test of ray tracing simulations in 
{\it UNI} case.

We also try a more sophisticated lens model using the same non-singular
finite isothermal ellipsoids with external shear. In this way the tidal
influence of masses close to the rays is at least partially represented. 

Models with included time delays information can be used to measure the 
Hubble's constant $H_0$ \citep{refsdal64}. We treat the value of the Hubble's 
constant as a free  parameter in some of our models, to check if it improves
the fits and (which is more important) to find the possible 
influence of {\it LOS}  and/or {\it ENV} on the accuracy of $H_0$ 
measurements.

The lens is described by six free model parameters ($q$, $\alpha_0$, $r_2$ 
defined in Sec.~2, the position angle in the sky, and two components 
of the lens position in the sky $\bm{\beta}_\mathrm{L}$; the other 
characteristic radius of the lens, $r_1$ is kept at constant ratio with 
$r_2$: $r_1=0.001~r_2$ and is not free). The source 
has some (unobserved) position in the sky $\bm{\beta}_\mathrm{S}$ 
(another 2 free parameters). Models with the external shear have another
2 free parameters ($\gamma_1$, $\gamma_2$), and the Hubble's constant 
$H_0$ may also serve as a free parameter. Thus the model has up to 
$N_\mathrm{par} = 8 + 2 + 1$ free parameters.

For a $N_\mathrm{im}$ image system we have $2N_\mathrm{im}$ observed 
positions and $N_\mathrm{im}-1$ independent flux ratios. If all relative 
time delays are measured, we have another $N_\mathrm{im}-1$ independent
observations. Including lens position in the sky we have 
$N_\mathrm{obs}=3N_\mathrm{im}+1$ (without time delays) or 
$N_\mathrm{obs}=4N_\mathrm{im}$ (with time delays) independent observations.
To have the positive number of the degrees of freedom 
$N_\mathrm{DOF}\equiv N_\mathrm{obs}-N_\mathrm{par}$ in all considered 
cases, we need $N_\mathrm{im}> 3$. In Paper I we considered all four and 
five image configurations. (The four image configurations, constituting 
about 5\% of all, were cases, where our numerical method failed to find 
the weakest central image.) Here we neglect the fifth (weakest) image 
(even if found) since it is usually not observed.

In this work we require that our models reproduce the image positions with
the accuracy of $\sigma_\mathrm{~I}=0.003~\mathrm{arcsec}$, the lens position
with $\sigma_\mathrm{~L}=0.1~\mathrm{arcsec}$, the image flux ratios 
(measured in stellar magnitudes) with $\sigma_\mathrm{~\Delta m}=0.1$, 
and time delays (when taken into account) with the accuracy of 
$\sigma_{~\mathrm{\Delta t}}=0.001~\mathrm{y}$. 
The rate of success of modeling depends on the chosen accuracy parameters 
(compare Paper I). The values we are using correspond to the typical errors 
of present day observations (e.g. \citealt{KZ04}), except for the  
high time delays accuracy, chosen for the reasons explained in Sec.~4.

We fit our models looking for a minimum of $\chi^2$. With the exception
of the term describing the errors in time delays modeling, we strictly follow 
the approach of Paper I. For completeness we repeat the formulae 
supplementing them with the time delay part: 
\begin{equation}
\chi^2=\chi^2_\mathrm{~L}+\chi^2_\mathrm{~I}+\chi^2_\mathrm{~\Delta m}
+\chi^2_\mathrm{~\Delta t}
\end{equation}
The first term, controlling the lens position, used in the model has 
the obvious form:
\begin{equation}
\chi^2_\mathrm{~L}=
\frac{(\bm{\beta}_\mathrm{L}-\bm{\beta}_\mathrm{L}^{~0})^2}
{\sigma_\mathrm{L}^2}
\end{equation}
where the subscript $\mathrm{L}$ stands for ``lens''.

Using the simplified model we can calculate the source positions 
$\bm{\beta}_\mathrm{S}^{~i}$ related 
to observed image positions 
$\bm{\beta}_\mathrm{I}^{~i0}$.
The deformation matrix $\mathsf{A}_{(i)}$ 
and magnification matrix $\mathsf{M}_{(i)}$ can be calculated at each 
image position:
\begin{equation}
\bm{\beta}_\mathrm{S}^i
\equiv\bm{\beta}_\mathrm{S}(\bm{\beta}_\mathrm{I}^{~i0}) 
~~~~~~~~
\mathsf{A}_{(i)}\equiv\left|\left|\frac{\partial\bm{\beta}_\mathrm{S}^{~i}}
{\partial\bm{\beta}_\mathrm{I}^{~i0}}\right|\right|
~~~~~~\mathsf{M}_{(i)}\equiv\mathsf{A}_{(i)}^{-1}
\end{equation} 
The mismatch between $\bm{\beta}_\mathrm{S}^{i}$ and 
$\bm{\beta}_\mathrm{S}$ 
implies magnified mismatch between modeled and observed image positions, 
which gives for $\chi^2_\mathrm{~I}$ (\citealp{K06}):
\begin{equation}
\chi^2_\mathrm{~I}=\sum_i~
\frac{\left|\mathsf{M}_{(i)}\mathbf{\cdot}
~\left(\bm{\beta}_\mathrm{S}^i-\bm{\beta}_\mathrm{S}\right)\right|^2}
{\sigma_\mathrm{~I}^2}
\end{equation}
The fitting statistic for flux ratios is given as:
\begin{equation}
\chi^2_\mathrm{~\Delta m}=\sum_{i=2}^{N_{im}}
\frac{\left(\Delta m_{i1}-\Delta m_{~i1}^0 \right)^2}
{\sigma_\mathrm{~\Delta m}^2}
\end{equation}
where we calculate the flux ratios relative to the first (brightest) 
image and express them in stellar magnitudes as $\Delta m_{i1}$. (The 
role of flux ratios is played by the ratios of corresponding lens 
magnifications.)  We then compare the modeled $\Delta m_{i1}$ and 
simulated $\Delta m_{~i1}^0$. 
For the time delays we use in full analogy:
\begin{equation}
\chi^2_\mathrm{~\Delta t}=
\sum_{i=2}^{N_\mathrm{im}}
\frac{\left(\Delta t_{i1} - \Delta t^0_{~i1}\right)^2}
{\sigma_\mathrm{~\Delta t}^2}
\label{chi2time}
\end{equation}
Again we calculate the time delays relative to the first (brightest)
image, comparing the modeled $\Delta t_{i1}$
and simulated $\Delta t_{~i1}^0$ differences of time propagation.
$N_\mathrm{im}$ is the number of images taken into account.

In all cases the quantities with an extra superscript ``$0$'' are taken 
from simulations and mimic the observed values, while 
the quantities without it result from modeling.

Following the approach of Paper I, we try to fit simulated image 
configurations with the simplified models. For the starting model we use 
the parameters of the lens used in the simulations. If the external shear
is included in modeling, we use its estimated value (see the next 
subsection). Standard minimization algorithm gives the best model. We accept 
it as a valid solution if its $\chi^2$ is within the range:
\begin{equation}
\chi^2 \le \chi^2_{~0.95}(\mathrm{DOF})
\end{equation}
where $\chi^2_{~0.95}(\mathrm{DOF})$ is taken from the table of $\chi^2$ 
distribution; statistically in 95\% of cases the valid models belong
to this range of $\chi^2$ for given DOF.

\begin{figure}
\includegraphics[width=84mm]{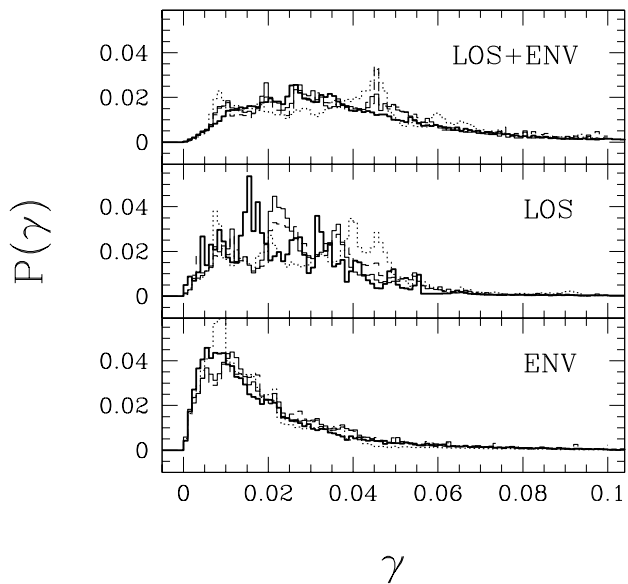}
\caption{Distribution of shear estimated with the first method (compare 
Eqs.~\ref{gam1},~\ref{gam2}) is shown with the thick solid lines. 
The distributions of the fitted shear values, 
are plotted with thin lines: dotted for model 4 (time delays neglected),
dashed for model 5 (time delays included, $H_0$ fixed), and solid for 
model 6 ($H_0$ treated as a free parameter).
In the upper panel the  influence of matter near the line of sight ({\it LOS}) 
and of the main lens environment ({\it ENV}) 
are taken into account; other panels show the effects separately.}
\label{shear}
\end{figure}

When simulating image configurations we use sources evenly spaced within 
caustic region of each lens. Thus every configuration represents some given 
solid angle of possible source positions $\Delta\Omega_l$. Its value is the 
same for all configurations related to a given lens $l$, but differs between 
the lenses. The magnification bias (\citealt{A81}; \citealt{P82}) makes the 
strongly amplified sources more likely to be included in the observed sample.
In general the effect depends on the shape of the source luminosity function; 
we adopt the simplification of \citet{KZ04} assuming power law form 
of this function with the power index $-2$, so the relative probability 
$p$ of including given image configuration to 
the sample is proportional to the total magnification $\mu$:
\begin{equation}
p = C\mu\Delta\Omega_l
\label{prob}
\end{equation}
where $C$ is a normalization constant.

Using the probability distribution we calculate the rates of acceptance of 
models of image configurations obtained using different approaches to light
propagation. The differences between individual lenses give the estimated
errors of the calculated rates. Our lens population is probably too 
small to be representative and the rates of acceptance can be used to compare 
the different approaches, but have no absolute meaning.
The results are shown in Table.~1. According to this table 
both {\it ENV} and {\it LOS} do have an impact on acceptability of fits and 
the latter seems more important. With the inclusion of the external shear
majority of the image configurations can be reproduced successfully. 
The time delays present an independent constraint on the solution
and lower the rate of acceptance of our models. 
The inclusion of $H_0$ as a free parameter slightly improves the
rate of acceptance of fits.

\subsection{External shear and convergence}

\begin{figure}
\includegraphics[width=84mm]{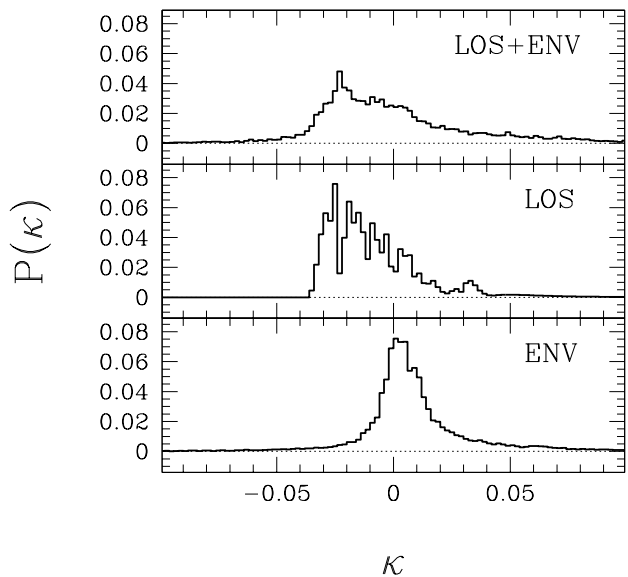}
\caption{Distribution of the estimated convergence ($\kappa$) due 
to the matter outside the main deflector (compare Eq.~\ref{kapp}).
The background contribution to the convergence is automatically 
included and is not shown separately. It is small 
($-0.007 < \kappa_\mathrm{bcg} < 0.008$) in all investigated cases. 
Naming conventions follow 
Fig.~1. }
\label{all_kappa}
\end{figure}

We estimate the external shear in two ways. In the first approach
the estimate of the external shear is done by removing the main 
lens and computing influence of its neighbours ({\it ENV}) 
and matter along the paths of the rays ({\it LOS}) in the weak lens 
approximation. After removing the main lens we calculate the deformation
matrix including only terms linear in deflection angles derivatives: 
$\mathsf{A}\equiv \mathsf{I}-\mathsf{H}$,
where $\mathsf{I}$ - is the identity matrix and $\mathsf{H}$ 
denotes the first order terms:
\begin{equation}
\mathsf{H} = \sum_{i=1}^{N-1}~\frac{d_{iN}}{d_N}~
\left|\left|\frac{\partial\bm{\alpha}_i}{\partial\bm{\beta}_i}\right|\right|
\label{matrix}
\end{equation}
With the usual convention we identify convergence and shear components:
\begin{eqnarray}
\kappa&=&\frac{1}{2}(\mathsf{H}_{11}+\mathsf{H}_{22})\label{kapp}\\
\gamma_1&=&\frac{1}{2}(\mathsf{H}_{11}-\mathsf{H}_{22})\label{gam1}\\
\gamma_2&=& \mathsf{H}_{12}\label{gam2}\\
\gamma&=&\sqrt{\gamma_1^2+\gamma_2^2}
\end{eqnarray}
Since we are interested in the mean shear acting in the region of the size 
similar to the main lens Einstein ring, we calculate the deflection angles 
derivatives in Eq.~\ref{matrix} using finite differencing with spacing
$\Delta\beta\approx\beta_\mathrm{E}$. 
We repeat shear calculations for many locations inside Einstein ring. 
We obtain the distributions of estimated shear due to {\it LOS} 
and {\it ENV} for each lens. We check the results with ten times 
smaller spacing obtaining very similar results. 
In Fig.~1 we compare the distribution of estimated shear with the 
distribution of its fitted values. This approach to shear 
estimation loosely resembles the cosmic shear measurement,  since in both 
cases the resulting influence on ray paths/image shapes is taken into 
account.

The method gives also the estimated values of the convergence. We are not 
using convergence in modeling, but we need it for the discussion 
of the results of Hubble's constant fitting. The distributions of 
the estimated convergence due to {\it ENV}, {\it LOS} and both are shown 
in Fig.~\ref{all_kappa}. 
The distributions are not Gaussian, so we characterize them by 
their {\it median} values ($\kappa^\mathrm{med}$) and the parameter
$\Delta\kappa$ such that 68\% of results belongs to the range
$\kappa^\mathrm{med}\pm\Delta\kappa$.
The numbers are: $0.005\pm 0.015$ ({\it ENV}), 
$-0.013\pm 0.016$ ({\it LOS}),
and $-0.007\pm 0.028$ ({\it LOS+ENV}) respectively. 
Our simulations include only eighty lenses and their environments 
belonging to eight independent small regions of space, so the convergence
distributions are not expected to be universal. In the {\it LOS} case 
the skewness and the range of $\kappa$ resemble the results of the 
recent study by \citet{SH11}, who investigate the influence of weak 
lensing on the spread in the observed signals from standard candles 
and standard sirens.

\begin{figure}
\includegraphics[width=84mm]{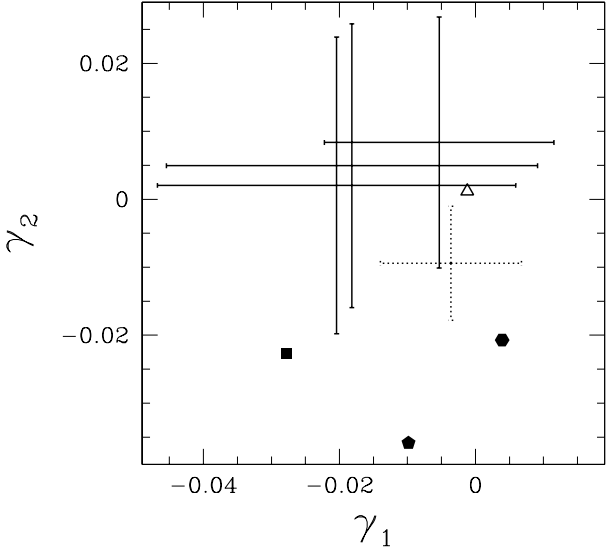}
\caption{The result of various methods of estimating the components $\gamma_1$ 
and $\gamma_2$ of the external shear are shown for one of the lenses as 
an example.
In all cases the {\it LOS+ENV} approach is used.  The solid error bars show
the shear values fitted with the help of models 4 -- 6. The spread of the 
results shows that the fitted shear value depends on the particular image 
configuration, not only on the lens and its environment.
Similarly the value of shear estimated by the first method 
(Eq.~\ref{gam1},~\ref{gam2}) 
is shown with the dotted error bar. In this case the estimated shear value 
depends on the position of the ray beams used in calculating the distortion 
matrix. The contribution from the background matter distribution (Sec. ~2.1)
is drawn as an open triangle. Finally the shear values estimated 
by the second method (Eqs.~\ref{sis1},~\ref{sis2}) taking into account 
external galaxies within the angular radius of $15$, $30$, 
and $60~\mathrm{arcsec}$ are shown with solid
square, pentagon, and hexagon respectively.}
\label{example}
\end{figure}

In the second approach, following \citet{Aug07}, we estimate the 
sources of shear taking into account contributions from
individual haloes, which are close to the lens on the sky. We use {\it SIS} 
model for each halo. The combined effect is given as:
\begin{eqnarray}
\gamma_1^{~\mathrm{ind}} = & 
\sum_{|\beta_i|<\beta_\mathrm{max}}~\frac{d_{i^\prime N}}{d_N}
~\frac{\alpha_i^{~\mathrm{SIS}}}{\beta_i}\cos(2\theta_i)
\label{sis1}\\
\gamma_2^{~\mathrm{ind}} = & 
\sum_{|\beta_i|<\beta_\mathrm{max}}~\frac{d_{i^\prime N}}{d_N}
~\frac{\alpha_i^{~\mathrm{SIS}}}{\beta_i}\sin(2\theta_i) 
\label{sis2}
\end{eqnarray}
We include halos within the circle of radius $\beta_\mathrm{max}$ around the 
main lens. The subscript $i$ enumerates the haloes and $i^\prime$ denotes 
the layer it belongs to, 
$\alpha^\mathrm{SIS}=2\pi v_\mathrm{vir}^2/c^2$ is the deflection
by a {\it SIS} halo with the virial velocity $v_\mathrm{vir}$, $\beta_i$ 
is the separation, and $\theta_i$ the position angle of the $i$-th halo 
relative to the lens.

The fits to image configurations do not reproduce the shear values 
estimated using the first approach exactly. The fitting gives results
which depend on the image configuration used in the particular optimization 
process and the estimates depend on the particular ray beams used in 
calculations. Treating fitting and estimating as two methods of measuring 
the shear value, we see that they are not in conflict: the distributions 
of fitted and estimated values are similar for combination of all lenses 
(compare Fig.~\ref{shear}) and the averaged values for a single lens 
obtained with the two methods are within standard deviation of each other 
(Fig.~\ref{example}). The shear values estimated by the second method
seem to be less reliable: comparing the influence of the lens neighbours 
on the sky within radii of $15$, $30$, and $60~\mathrm{arcsec}$ we see 
the dependence of the result on the size of the region taken into account.
Our maps are too small, to check still larger regions (even the 
$60~\mathrm{arcsec}$ radius is not applicable to all our lenses), however 
the dependence seen in the results suggests that the saturation expected in 
the limit of large radii has not been reached. The analysis of fitted 
and inferred shear values for a sample of real quad lenses is presented by 
\citet{Won11}. According to their study the values of the lens model shears 
and environment shears are in general different and in three of six 
investigated cases they are inconsistent at greater than 95\% confidence. 

\begin{figure}
\includegraphics[width=84mm]{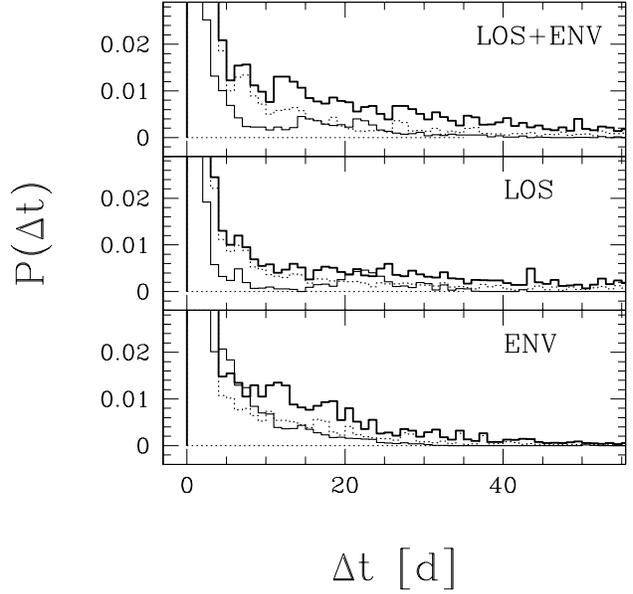}
\caption{Distribution of the errors in time delays ($\Delta t$).
Initial errors for all configurations are shown with 
thick solid lines; the sum of all bins gives one for these histograms.
The distributions of time delay errors for acceptable models including shear
but neglecting time delays are shown with thin lines;
these histograms are normalized by the rate of success (column 4 in 
Table 1). The dotted lines show the distributions before fitting, 
the thin solid lines - after fitting image positions and flux ratios.
The subsequent panels show results for the different 
approaches to light propagation, using the convention of  Fig.~1.}
\label{delays}
\end{figure}

\subsection{Simulated and fitted time delays}

We define
\begin{equation}
\Delta t = \sqrt{\frac{\sum_{i=2}^{N_\mathrm{im}}
\left(\Delta t_{i1} - \Delta t^0_{~i1}\right)^2}{N_{im}-1}}
\end{equation}
as a measure of the error in model time delays.

In Fig.~\ref{delays} we present the probability distribution for the 
values of the time delays error. Our aim is to find whether successful 
fitting of image positions and flux ratios with a simplified model
automatically improves the time delay errors. The thick solid lines in 
Fig.~\ref{delays} show the distributions of initial time delay errors
for all configurations and dotted lines - for configurations successfully 
fitted with models neglecting time delays. These initial values are 
calculated for models having the parameters of the lens used in 
simulations and for the expected value of the external shear. 
The thin solid lines show the distributions of time delay errors after 
the fit. Comparing the plots we see that a small initial error in time 
delays gives a better chance of successful fit. Fitting of image positions 
redistributes the time delay errors. Examining the numerical data we find,
that fitting the image positions increases the fraction of models with
acceptable time delay errors ($\Delta t < 1^\mathrm{d}$) by 10 -- 20 per 
cent, but another 15 -- 25 per cent still have $\Delta t$ too large. 
(Large fraction of these, but not all, can still be improved with 
fitting procedure using time delays). Our results show that in $\sim 2/3$ -- 
$4/5$ of cases reproducing image positions and flux ratios implies also
reproducing time delays with good accuracy.

\subsection{The Hubble's constant}

The value of the Hubble's constant $H_0$ defines the length unit 
in the Universe, $c/H_0$, and the time unit, $1/H_0$. Scaling the distances
 and sizes of all mass concentrations, but preserving their velocity 
dispersions, positions, orientations, and shapes, we get the same lens 
equation and the same image configurations. Since relative time delays
scale as well, fits using time delays can be used to measure the scaling 
factor and so the Hubble's constant \citep{refsdal64}.

\begin{figure}
\includegraphics[width=84mm]{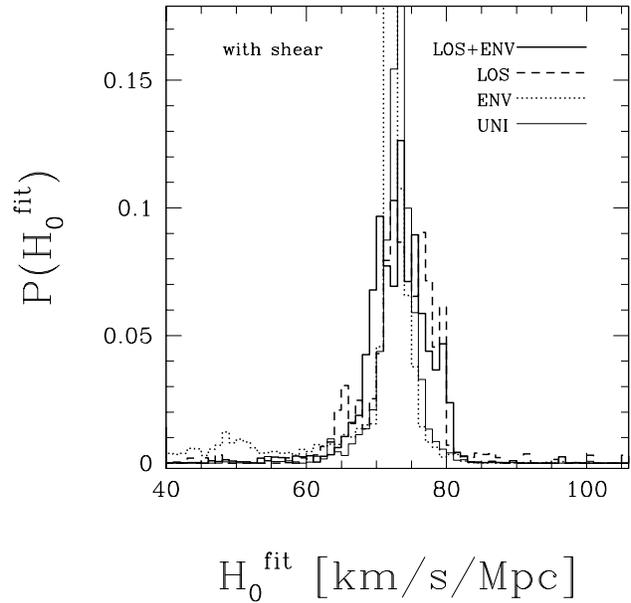}
\caption{Probability distribution for the fitted values of Hubble's constant
based on configurations including the external shear in modeling. 
The simulations have been performed in a universe model with 
$H_0=73~\mathrm{km/s/Mpc}$. The histograms show the results only for the 
acceptable  fits. The distribution of fitted $H_0$ values depends 
on the approach to light propagation (see the 
text for details). The plotting conventions are explained within the figure.}
\label{hub_yes}
\end{figure}

Using our lens models with $H_0$ treated as a free parameter and applying 
it to all quad image configurations we obtain distributions of estimated 
Hubble's constant values corresponding to different descriptions of light 
propagation, using different lenses and different image configurations. 
The {\it UNI} case shows the typical accuracy of our method. Despite the fact 
that an initial model has parameter values the same as used in 
simulations and is acceptable, the fitting program looks for a minimum 
of $\chi^2$ which usually lies close to but not exactly at the initial
position. This affects all the fitted parameters. 

The distributions of fitted Hubble's constant values are not symmetric, 
with outliers at positions far from $H_0$ used in simulations, so we 
characterize them using median values $H_0^\mathrm{med}$ and absolute 
deviation $\Delta H_0$ such, that 68\% of results belong to the 
$H_0^\mathrm{med}\pm\Delta H_0$ range. The distributions of 
$H_0^\mathrm{fit}$ are shown in Fig.~\ref{hub_yes} for models
including the external shear (model 6 in Table~1).

The results of the Hubble's constant fitting depend to some extent 
on the way the external shear and convergence are treated. In Table~2
we present the median values and absolute deviations of the fitted Hubble's 
constant for models neglecting the external shear (column 1) and models 
taking it into account (column 2). We also check the distributions of the 
corrected $H_0$ value (see \citealp{KZ04}, \citealp{gfs88})
\begin{equation}
H_0^\mathrm{cor} = (1-\kappa)H_0^\mathrm{fit}
\label{corected}
\end{equation}
where the convergence $\kappa$ is estimated using Eq.~\ref{kapp} and for 
each lens its averaged value is employed. We do not introduce the convergence 
into our models but only correct the results of fitting according to 
Eq.~\ref{corected}.

\begin{table}
 \centering
 \begin{minipage}{75mm}

  \caption{Median values and their deviations for the fitted Hubble's constant [km/s/Mpc]}
  \begin{tabular}{@{}cccc@{}}
  \hline
         	& neglected & fitted & corrected \\
 \hline
 {\it LOS+ENV} & 70.5$\pm$ 5.3 & 72.6 $\pm$ 4.2 & 72.9 $\pm$ 3.5 \\ 
 {\it LOS}     & 70.5$\pm$ 6.4 & 73.1 $\pm$ 5.0 & 73.4 $\pm$ 4.1 \\
 {\it ENV}     & 72.3$\pm$ 1.8 & 71.5 $\pm$ 2.7 & 71.2 $\pm$ 2.7 \\
 {\it UNI}     & 72.8$\pm$ 0.8 & 72.8 $\pm$ 1.1 & 72.8 $\pm$ 1.1 \\
 \hline

\noalign{\vskip3pt}
\multicolumn{4}{p{7.5cm}}{Note: The table shows the dependence of 
the fitted median value and its absolute deviation
($H_0^\mathrm{med}\pm\Delta H_0$) of the Hubble's constant
on the method of treating the light propagation in simulations 
and on the way the external shear and convergence are treated.
The cases with zero shear and convergence (``neglected''), 
the shear treated as a free parameter and no convergence (``fitted''),
and the fitted shear with $H_0$ corrected according to Eq.~\ref{corected} 
(``corrected'')  are presented in consecutive columns. Naming of rows 
follows Table~1.
}
\end{tabular}
\end{minipage}
\label{hubvalues}
\end{table}

Statistically the fitted $H_0$ values are in agreement with their 
expected values. In the most realistic {\it LOS+ENV} case 68\% of the 
results, after the correction  using Eq.~\ref{corected} have errors below 5\%.
The median estimated convergences are close to zero; the {\it LOS}
case with $\kappa^\mathrm{med}=-0.013$ shows the largest departure.
For {\it LOS+ENV} and {\it LOS} cases the inclusion of the external shear 
improves the  median fitted $H_0$ and its accuracy. The correction improves 
the accuracy in both cases and the median in the first case. For the 
{\it ENV} case the opposite is true. The convergence we are using 
is estimated with the method of subsection 3.2, which is not perfect
(compare shear estimated and fitted values).  This method cannot be 
applied to real lenses since it is not based on observed properties
of the image configuration.

In our statistical approach we have neither studied the individual 
lenses, nor the individual image configurations. Some experiments suggest 
that the constraints on the $H_0$ value from a single configuration are 
rather poor. For many image configurations values of $H_0^\mathrm{fit}$ 
from a wide range may give fits of comparable quality.

\subsection{Lens parameters}

Modeling simulated image configurations with simplified models
we get the lens parameters. Since we
know the ``true'' parameters of the lens used in rays tracing, we are 
able to estimate the systematic errors introduced by {\it LOS} and {\it ENV}
effects. The results are presented in the following figures. 
We investigate the lens axis ratio $q$, 
and the characteristic deflection angle $\alpha_0$. 
With four image configurations the lens virial radius is not sensibly 
constrained and the same can be said of the lens virial mass 
defined in Eq.~(\ref{virialmass}). The knowledge of fitted $q$ and $\alpha_0$ 
allows for a good estimate of mass within a given radius.
 
\begin{figure}
\includegraphics[width=84mm]{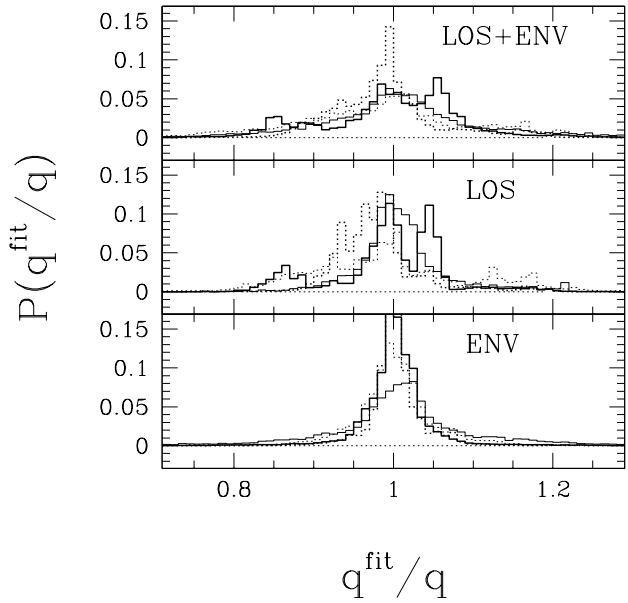}
\caption{Probability distribution for the values of the lens axis 
ratio. Acceptable models with external shear and using time delays 
(model 5 in Table~1) 
are shown with thick solid lines, with external shear but no time delays (4) - 
with thin solid lines, without shear and with time delays (2)- with thick 
dotted lines, and without shear or time delays (1) - with thin
dotted lines. The models treating $H_0$ as a free parameter (not shown)
give results very similar to models including time delays.
The subsequent panels show results for the different 
approaches to light propagation, the convention follows Fig.~1.}
\label{q-ratio}
\end{figure}

\begin{figure}
\includegraphics[width=84mm]{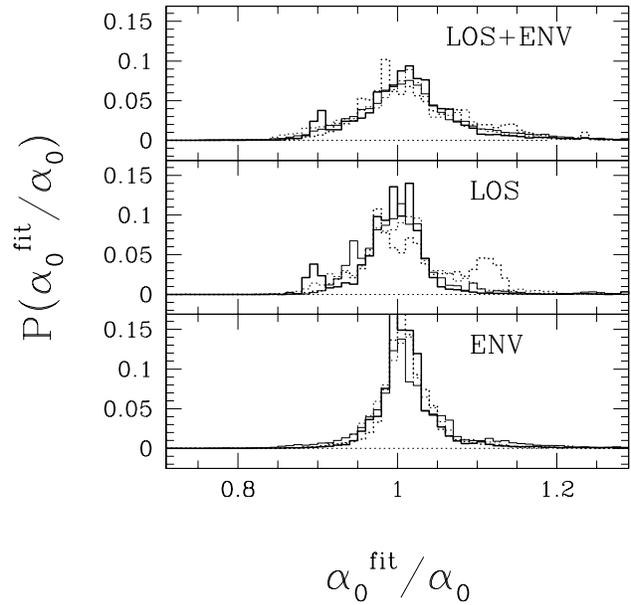}
\caption{Probability distribution for the ratios of the fitted to original 
values of the characteristic deflection angle. All conventions follow 
Fig.~\ref{q-ratio}.}
\label{alfa}
\end{figure}

For every parameter we find the ratio of its fitted (denoted by the 
superscript ``fit'') to original value. 
The dotted lines in Figs.~\ref{q-ratio} -- \ref{alfa} 
denote the results for the models, which 
do not use the external shear as a parameter, the solid lines - 
for models using it.

\begin{figure}
\includegraphics[width=84mm]{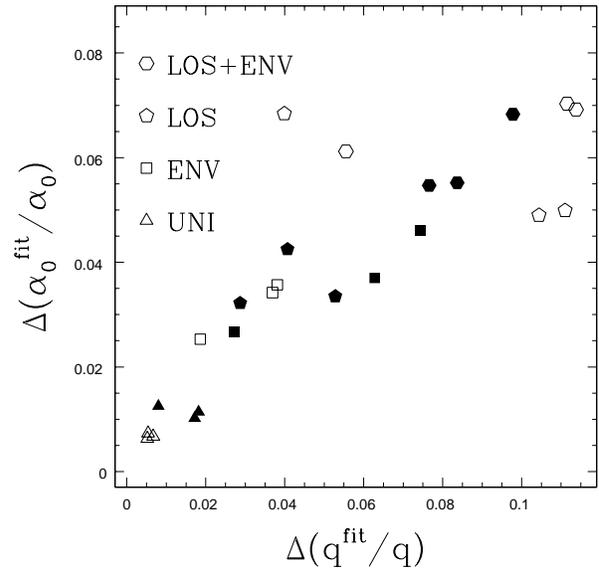}
\caption{The absolute deviation of the fitted values of axis 
ratio versus the absolute deviation of the fitted characteristic 
deflection angle. 
The open polygons represent models neglecting 
the external shear (models 1 -- 3 in Table~1), while solid polygons - 
models using it in in the fits (4 -- 6). 
The shapes correspond to different approaches to light propagation: 
triangles  -- {\it UNI},
squares  -- {\it ENV},
pentagons  -- {\it LOS},
and hexagons  -- {\it LOS+ENV}. }
\label{madpar}
\end{figure}
In Fig.~\ref{q-ratio}
we show the distributions of the ratios $q^\mathrm{fit}/q$ for models 
including or not the external shear and time delay data, and for 
different approaches to light propagation. (See the caption). As can be 
seen in this figure (and also in Fig.~\ref{alfa}), 
the distributions are rather complicated and depend both on the 
lens model and the approach to light propagation. Figs.~\ref{q-ratio} and 
~\ref{alfa} suggest that the widths of the distributions increase 
systematically when going through the {\it ENV}, {\it LOS}, and 
{\it LOS+ENV} cases, but the absolute deviations plotted in Fig.~\ref{madpar}
do not fully support this simplified view. The median $q^\mathrm{fit}/q$ 
value is always close to unity (with 1\% accuracy).

In Fig.~\ref{alfa} we show the distribution of the ratios 
$\alpha_0^{~\mathrm{fit}}/\alpha_0$ for the characteristic 
lens deflection angle. 
The median $\alpha_0^{~\mathrm{fit}}/\alpha_0$ value 
is always close to unity (1\% accuracy).
The characteristic deflection angle is related to the velocity dispersion
of stars in the lensing galaxy in the same way as in simpler {\it SIS} lens 
models, i.e. $\alpha_0 \sim \sigma^2$.  Since 
$\Delta(\alpha_0^\mathrm{fit}/\alpha_0)<0.08$ (compare Fig.~\ref{madpar}), 
the velocity dispersion derived from the lens modeling has typical 
relative error $\sim 0.04$.

We also compare the influence of {\it LOS} and {\it ENV} on the 
errors in the fitted parameter values showing the absolute 
deviations of the fitted $q^\mathrm{fit}/q$ and 
$\alpha_0^\mathrm{fit}/\alpha_0$ ratios 
in one diagram. As can be seen in Fig.~\ref{madpar}, the typical errors 
in both parameters are roughly proportional to each other.

\section{Discussion and conclusions}

In this paper we have followed the approach of Paper I, investigating the 
influence of the matter along the line of sight on the properties of  
strong gravitational lenses. The main difference is the calculation of 
propagation time in the simulations and inclusion of the relative time 
delays in the modeling, which may serve the analysis of the Hubble's 
constant measurement.

As compared to Paper I, we have simplified the treatment of the background
matter distribution (compare Sec.~2.1), effectively replacing a part 
of the numerical calculation by the use of the fitted analytical formulae.
This approach was not employed in Paper I,  so the numerical 
inconsistencies, measured by the rotation to shear ratio (up to $\sim 0.1$)
were more important and could have affected its results, exaggerating 
the reported influence of the matter along the line of sight on strong 
lensing.

Our numerical experiments show that matter in the immediate lens vicinity
and matter along the line of sight do influence image configurations 
and values of fitted lens parameters . If the Hubble's constant is treated 
as a free parameter, its fitted value also depends on the matter outside 
the lens. These conclusions are not new (see e.g. \citealp{KZ04}) and have 
been demonstrated to be true in several cases of real lenses 
(\citealp{Will06}, \citealp{Mom06}, \citealp{Aug07}). According to our 
results matter along the line of sight is more important than the 
lens immediate neighbours. This may cause problems in modeling multiple 
images and time delays of high redshift sources, where direct observations 
of the matter along the line of sight may be impossible.

We consider a ``typical'' strong lens and its environment using a synthetic 
model of the matter distribution based on the cosmological Millennium 
Simulation \citep{Spr05} with gravitationally bound haloes described in
\citet{b11} and \citet{b4}. We investigate eight randomly chosen small 
square regions on the synthetic sky (each $\approx 3~\mathrm{arcmin}$ 
on a side) and look for the ten strongest lens candidates in each region,
assuming that a source is located somewhere in the background at the redshift
$z\approx 2$. We artificially remove the lens neighbours and/or matter 
inhomogeneities along the line of sight to compare the resulting image 
configurations. Thus we obtain four different matter distributions with 
the same strong lens. To assess the importance of matter outside the lens 
we try to model the image configurations using a single lens in a uniform 
universe model. The rate of failure of this approach, as well as the spread
of the fitted lens parameters serves as a measure of the influence 
of the matter outside the lens on its properties. Inspecting Table.~1 
one can see that (statistically) matter along the line of sight 
is more important as compared to the lens close neighbours. The same 
conclusion follows from inspecting Figs.~\ref{q-ratio} -- \ref{alfa}, 
which show the distributions of fitted lens parameters.

The inclusion of the time delays in modeling lowers the rate of 
acceptance of fits not more than 10\%, despite the high postulated  
time measurements accuracy ($0.001~\mathrm{y}$). We have introduced this 
accuracy requirement because many of our image configurations give time 
delays of the order of couple of days. For them the time delay modeling 
and Hubble's constant fitting with more realistic $\sim 2^\mathrm{d}$ accuracy 
would be meaningless. The time delay measurement accuracy 
we are using in our calculations is probably beyond the present 
observational possibilities. On the other hand one can choose appropriate 
lenses with appropriate image configurations for the purpose of 
estimating $H_0$. This however goes beyond the scope of this paper, where 
we examine a randomly chosen lens population as a whole. 

Treating the value of the Hubble's constant as a free parameter and using
the simplified lens model with the external shear we get its median 
fitted value in agreement with the ``true'' value used in simulations 
(compare Table 2 and Fig.~\ref{hub_yes} in Sec.~3.4).  
The convergence caused by matter outside the lens, when averaged 
over all our investigated cases, has median value close to zero (Sec.~3.2), 
so  the bias in fitted $H_0$ values should not be pronounced.

The expected convergence along a randomly chosen line of sight 
is zero since we measure it relative to the uniform universe model.
The haloes are compensated, so for each of them the above statement 
is also true. Since the positions of haloes in space are correlated, 
the density of matter near any lens should be on average increased. 
On the other hand, when calculating the convergence we remove the 
contribution of  the lens itself, which is dominant. Thus the external
convergence value close to zero on average, as in the case of lens 
population investigated here, is not implausible. In a more realistic 
approach the convergence near the main lens should be positive due
to the correlations in matter distribution. Simply increasing the radii 
of compensating disks would not help, 
since the influence  of each individual halo would diminish, 
but the number of haloes taken into account would increase 
in inverse proportion. One may introduce some artificial limiting radius
independent of masses and take into account only lenses within such limits.
We avoid such approach since the influence of high mass, distant haloes 
would be lost and the compensation would not be complete. 
(A different method of compensating mass of haloes was 
presented in \citealp{JK10}).

Some of the lenses give fitted Hubble's constant values far from the 
median. We have checked the accuracy of our method applying it to lenses 
surrounded by the 
uniform matter distribution and found that it gives $\sim 2$ per cent 
scatter in fitted $H_0$. The outliers cannot be explained as a result of 
unusually large convergence. Probably the perturbations to the outliers 
are nonlinear and cannot be modeled as the external shear and convergence.

Our calculations show that the external shear used in simplified modeling 
can be roughly estimated by weak lensing measurements along the line of 
sight 
(compare Figs.~\ref{shear},~\ref{example} and  Eqs.~\ref{gam1},~\ref{gam2}).
On the other hand taking into account only galaxies visible on the sky close 
to the lens position
(Fig.~\ref{example}, Eqs.~\ref{sis1},~\ref{sis2}) may be insufficient 
-- we have checked such estimates using lens neighbours on the sky in 
circles of radii up to $\sim 1~\mathrm{arcmin}$ and found the dependence 
of the results on the size of the region. It shows 
the important role played by the large scale structure, which is difficult 
to include in modeling. More sophisticated methods using spectroscopic 
observations of galaxies along the line of sight (\citealp{Mom06}, 
\citealp{Aug07}) may give better estimates of the external shear.

In this paper we have limited our interest to configurations of four images. 
In five image configurations we neglect the fifth, weakest image, which is 
not observed in real lenses.  (This is another difference in 
our approach as compared to Paper I.) 
Neglecting the fifth image we loose the information related to the lens centre.
The nonsingular isothermal ellipsoids we are using have small,
approximately constant surface density cores. The size of the core 
has impact on the fifth image brightness, so without the fifth image 
it cannot be fitted. In simulation we postulate that the core size 
is proportional to the lens virial radius, which defines the total mass
of the lens. Again, without the fifth image the total mass cannot be fitted.
Of course the mass inside the Einstein ring can still be estimated using the 
fitted value of characteristic deflection angle $\alpha_0^{~\mathrm{fit}}$.

\section*{Acknowledgments}
We are grateful to the Anonymous Referee, whose critical remarks greatly 
improved the paper.
The Millennium Simulation databases used in this paper
and the web application providing on-line access to them were constructed
as part of the activities of the German Astrophysical Virtual Observatory.
We are grateful to Volker Springel for providing us with the smoothed
Millennium density distribution in the early stage of this project. 
This work has been supported in part by the Polish 
National Science Centre grant N N203 581540.

\end{document}